\documentclass[twoside,onecolumn]{article}
\usepackage{blindtext} 

\usepackage[sc]{mathpazo} 
\usepackage[T1]{fontenc} 
\usepackage{microtype} 

\usepackage[english]{babel} 

\usepackage[hmarginratio=1:1,top=32mm,columnsep=20pt]{geometry} 
\usepackage[hang, small,labelfont=bf,up,textfont=it,up]{caption} 
\usepackage{booktabs} 

\usepackage{lettrine} 
\usepackage{jinstpub}
\usepackage{enumitem} 
\usepackage[T1]{fontenc}
\usepackage[utf8]{inputenc}
\usepackage{comment}
\usepackage{epsf}  
\usepackage{epsfig}  
\usepackage{graphicx}
\usepackage{amsmath}
\usepackage{tabularx}
\usepackage{xspace}
\usepackage{float}
\usepackage{esvect}
\usepackage{morefloats}
\usepackage{subfloat}
\usepackage{subfig}
\usepackage{relsize}
\usepackage[english]{babel}
\usepackage[babel]{csquotes}
\usepackage{caption}
\usepackage{hyperref}
\usepackage[left]{lineno}
\usepackage{rotating}
\usepackage{mathtools}
\usepackage{shortcuts}
\usepackage{siunitx}
\usepackage{multirow}
\usepackage{fixltx2e}
\usepackage{amsmath}

\usepackage{titling} 

\usepackage{hyperref} 
\usepackage{authblk}

\title{A facility for radiation hardness studies based on the Bern medical cyclotron} 
\author[a]{ J. Anders}
\author[a]{S. Braccini} 
\author[a]{T. Carzaniga} 
\author[a]{A. Ereditato}
\author[a]{A. Fehr}
\author[a]{F. Meloni}
\author[a]{C. Merlassino}
\author[a]{A. Miucci}
\author[a]{M. Rimoldi}
\author[a]{M. Weber}
\affil[a]{ AEC/LHEP, Universit\"at Bern } 

\date{\today} 
\renewcommand{\maketitlehookd}{%
\begin{abstract}
\noindent The development of instrumentation to be operated in high-radiation environments is one of the main challenges in fundamental research. Besides space and nuclear applications, particle physics experiments also need radiation-hard devices. The focus of this paper is a new irradiation facility based on the medical cyclotron located at the Bern University Hospital (Insespital), which is used as a controlled $18\,\MeV$ proton source. The adjustable beam current allows for dose rate dependent characterisation over a large dynamic range, from 0.1 to 1000\,\si{\giga\radian} per hour. The beam can be tuned so that the user can obtain the desired irradiation conditions. 
A complete study of the device under irradiation is possible thanks to dedicated beam monitoring systems as well as a power control system for the device under irradiation, which can be operated on-line. Further characterisations of the irradiated devices are possible thanks to a laboratory equipped with gamma spectroscopy detectors, ammeters and transient current technique setups. 
\end{abstract}
}

\begin{document}

\maketitle


\section{Introduction}

A medical cyclotron is in operation at the Bern University Hospital (Inselspital). It is a commercial IBA18/18 cyclotron, accelerating $H^-$ to $18\,\MeV$\,\cite{Cyclo}. The machine is equipped with an external $6.5\,\si{\meter}$ long beam transfer line which delivers the beam in a second bunker, referred as irradiation bunker in this paper. This feature makes it possible to conduct routine radio isotope production for PET diagnostics in parallel with multidisciplinary research activities. The Bern Cyclotron can be operated for research activities during the day time while radio isotope production for medical purpose is running over-night.
The research program at the Bern Cyclotron includes the production and characterisation of radio-isotopes, the development of beam-monitoring detectors, and the characterisation of radiation-tolerant devices. The last issue is the subject of this paper.
Radiation tolerant devices are fundamental in high-energy physics, space, nuclear and medical applications. In order to use a medical cyclotron for irradiation purposes a tuning and characterisation campaign was done, in particular the beam energy profile was measured and the cyclotron was tuned for low output current operation. It is fundamental to have an accurate dose measurement so that an beam monitoring setup was developed to provide a precise measurement of the ouput current. As a results of these studies the Bern Cyclotron allows for an accurate adjustment of the beam current, in this way it is possible to make real-time studies on the device under irradiation. It is hence possible to obtain different radiation conditions or allow dose rate dependent characterisation. The Bern Cyclotron allows for radiation hardness studies with a lower limit of the dose at $\sim\,10\,\si{\kilo\radian}$.
The facility is fully equipped to host custom readout of the device during the irradiation. Temperature control of the device is possible with Peltier elements and/or water cooling.
A dedicated laboratory is available on-site equipped with germanium and scintillating detectors for measuring the samples emission spectrum and activity. A high performance ammeter for electrical measurement is also available. Further characterisation is possible in the nearby Laboratory for High energy Physics (LHEP) thanks to transient current technique setups, FPGA custom logic setups for electronic device characterisation and high-voltage power supplies for the current-voltage characterisation.

\section{The Bern cyclotron and its beam transport line}

\begin{figure}
	\centering

		 \includegraphics[width=0.8\textwidth]{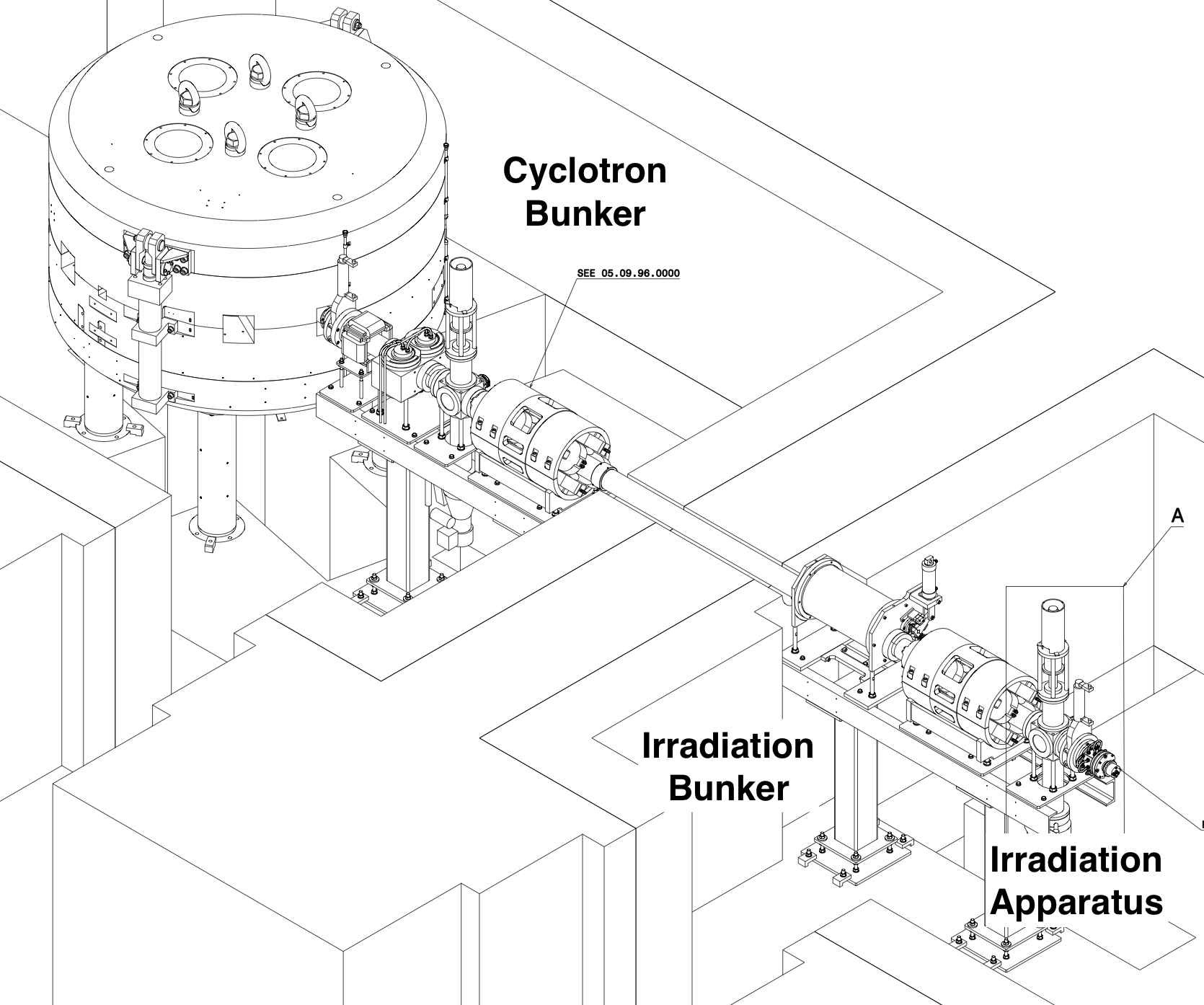}
		 \caption{Diagram of the cyclotron facility, the cyclotron bunker is separated from the irradiation bunker where the irradiation apparatus sits. This facilitates the access to the device under irradiation due to the low level of activity in the irradiation bunker, below $2\,\si{\micro\sievert}$ per hour.}
		 \label{fig:Cyclo_scheme}

\end{figure}

The Bern Cyclotron is an IBA Cyclone $18\,\MeV$ cyclotron which is equipped with two $H^-$ sources. It can provide beam currents up to $150\,\si{\micro\ampere}$ in single or dual beam modes. This current is never reached in a typical irradiation and it is limited to few hundred $\si{\nano\ampere}$ output. Proton extraction is achieved by stripping off the electrons from the negative $H^-$ ions with $5\,\si{\micro\meter}$ thick pyrolytic carbon foils. A detailed description of the Bern cyclotron can be found in Ref.\,~\cite{Cyclo}.
The beam is transported to the irradiation bunker with an transfer efficiency of greater than $95\%$ via a $6.5\,\si{\meter}$ long beam transfer line (BTL). The research bunker has a low level of activity, below $2\,\si{\micro\sievert}$ per hour, allowing access to the beam area. The irradiation bunker can be then accessed a few minutes after the beam is switched off, making intervention possible. For this reason an irradiation facility was setup there as shown in Figure \ref{fig:Cyclo_scheme}. The accelerated protons are steered by an X-Y magnet system and are directed tEq.owards a collimator. The focusing of the beam is realised using two horizontal-vertical quadrupole doublets; the first is located in the cyclotron bunker and the second in the irradiation bunker. A $1.8\,\si{\meter}$ thick wall divides the bunkers and is necessary to stop the fast neutron flux generated during isotope production. For the same reason, a movable cylindrical neutron shutter is located inside the beam pipe while the beam transfer line is not in use. Two beam viewers are present in the line allowing for beam current measurements.
Despite the design of the cyclotron for high current operation (with maximum of $150\,\si{\micro\ampere}$), operation with currents at the $\si{\pico\ampere}$ level is possible. A dedicated tuning was performed for this purpose. This is not a standard setting for commercial cyclotrons and, as a result, additional developments were required. A detailed description of the tuning process can be found in Ref.\,\cite{Cyclo_LC}. Optimised settings were obtained for the operation of the ion source, the main coil, the radio-frequency systems and for the BTL magnets.

\section{The dose measurement method}

The measurement of the dose is essential to link the evolution of the device performance to the bulk and surface damages caused by the radiation. A proton beam interacts with the material by means of ionisation and nuclear processes. It is important to measure the absorbed dose due to the two different components, which are both proportional to the flux of protons hitting the device.
The dose due to ionisation ($\mathcal{D}_{ion}$) and nuclear ($\mathcal{D}_{nuc}$) processes can be calculated with Eq. \ref{eq:TID_est}:
\begin{align}
\centering
&\mathcal{D} = \phi \cdot \dfrac{1}{\rho}\big( \dfrac{dE}{ dx}_{ion} + \dfrac{dE}{dx}_{nuc} \big) = \mathcal{D}_{ion} + \mathcal{D}_{nucl};  \, \label{eq:TID_est} 
\end{align}
where $\phi$ is the fluence, defined as number of incident particles per unit area, $\rho$ the density of the target and $\dfrac{dE}{dx}_{ion/nuc}$ is the average energy loss per unit of length in a given material for either ionisation or nuclear interactions. 
The effects of nuclear interaction are often compared to a reference 1$\MeV$ neutron fluence as can be seen in Eq.\,\ref{eq:Neq}
\begin{align}
\centering
&\phi(1\,\MeV\quad n_{eq}) =\dfrac{\int_0^{+\infty} D(E) \phi(E)dE}{D(\rm{1MeV}\quad n_{eq})}. \label{eq:Neq}
\end{align}
where $\phi(E)$ is the fluence of certain kind of particles at a given energy E and D is the damage function which depends upon the target material, the characteristics of the incident particle (energy, mass and electrical charge).
The measurement of the fluence is used to assess the dose, for both nuclear and ionising phenomena, if the proton energy and the material properties are known.
The fluence can be obtained by measuring the beam current:
\begin{align}
\centering
\phi &= \int _{t_0} ^{t_1}\frac {j(t)}{Q_{e}} dt; \, \label{eq:fluence}
\end{align}
where $j(t)$ is the current per unit area, $Q_e$ is the charge of the proton and $[t_0,t_1]$ is the time interval of the irradiation. A measurement of the current per unit area during the irradiation allows to measure the dose rate and at the same time to have a complete history of the irradiation. In particular, the dose rate measurement is relevant for identyfing non-linear phenomena which are typical in electronics, as presented in Ref.\,\cite{FEI4_irr}.

\section{The irradiation setup}
\begin{figure}
	\centering
	\includegraphics[width=\textwidth]{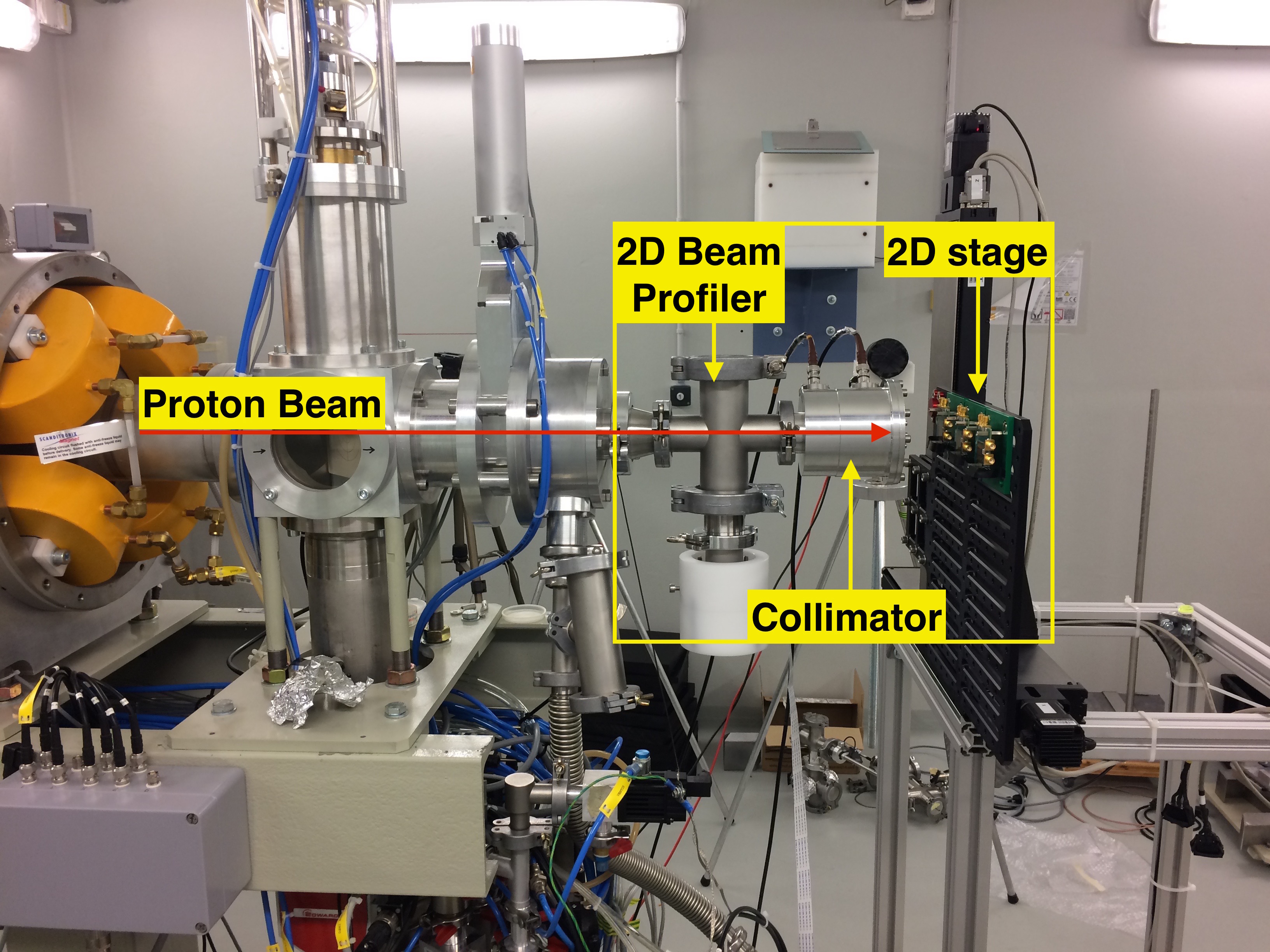}
	\caption{Irradiation setup at the Bern Cyclotron. It consists of the two-dimensional beam profiler, for measuring the beam transverse profile, and the the collimator system for the current measurement and the beam shaping. The 2D stage, for alignment of the devices under irradiation, is visible on the right. }
	\label{fig:irr-fac}
\end{figure}

A dedicated irradiation setup was developed for the measurement of $j(t)$. The method relies on the extraction of the central part of the beam to irradiate the device, while dumping the outermost part of the beam for measuring the current density. 
It is important to operate the cyclotron so that the delivered beam has a constant intensity per unit surface in the transverse plane. The transverse uniformity guarantees a uniform irradiation of the device and as well as a constant beam density over the dump area. In order to obtain the required beam profile, the downstream quadrupole doublets of the BTL are switched off. In this way the beam profile is flat to within $20\%$ up to $40\,\si{\milli\meter}$ in both the vertical and the horizontal directions.
The irradiation setup is shown in Fig.\,\ref{fig:irr-fac}. It is composed of two detectors: a two-dimensional beam profile detector and a collimator to measure the density current and to extract the beam with a custom profile. The extraction window consists in a $300\,\si{\micro\meter}$ thick aluminum layer. A 2D movable stage completes the system, allowing for the irradiation of large surface devices, up to $30\times30\,\si{\centi\meter}^2$, or the irradiation of several devices without accessing the irradiation bunker. The 2D-stage has a $10\,\si{\micro\meter}$ precision, more details can be found in Ref.\,\cite{intellidrives}.

\begin{figure}
	\centering
	{\includegraphics[width=0.8\textwidth]{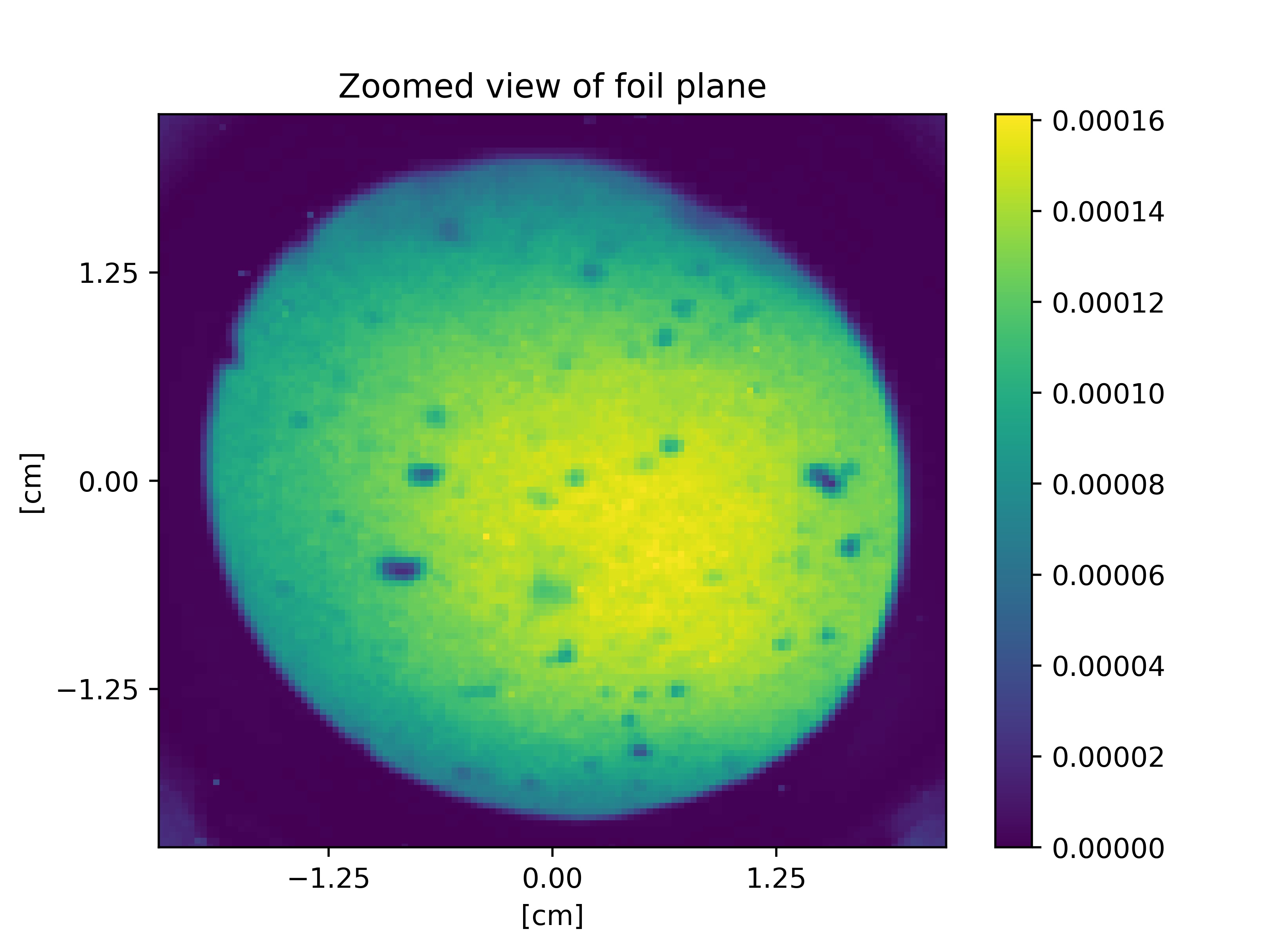}}
  	\caption{Beam shape obtained with the two-dimensional beam profiler. The colour scale is in arbitrary units. The dark dots in the picture are due to a poor ytterbium deposition on the device. The beam is flat to within 20\%. A correction factor $\Lambda$ is derived to take in account this non-homogeneous transverse profile.}
	\label{fig:2D_profile}\quad

\end{figure}

The two-dimensional beam profiler is a prototype detector developed at University of Bern which consists of a matrix of silicon dioxide enriched with ytterbium \cite{schefer2018,schmid2018}. Protons hitting the matrix produce fluorescence in ytterbium. The matrix is placed at an angle of $45\,\si{\degree}$ to the beam and the fluorescence is monitored by a National Instrument Smart Camera\,\cite{NIM_smartCamera}. Further processing is performed to obtain the final bi-dimensional and one-dimensional profiles. A 2d image of the beam is shown in Fig.\,\ref{fig:2D_profile}. A paper describing the detector is in preparation. The system was calibrated and aligned with the UniBeam detector\,\cite{UniBeam}. 



\begin{figure}
	\centering
				 \includegraphics[width=0.9\textwidth]{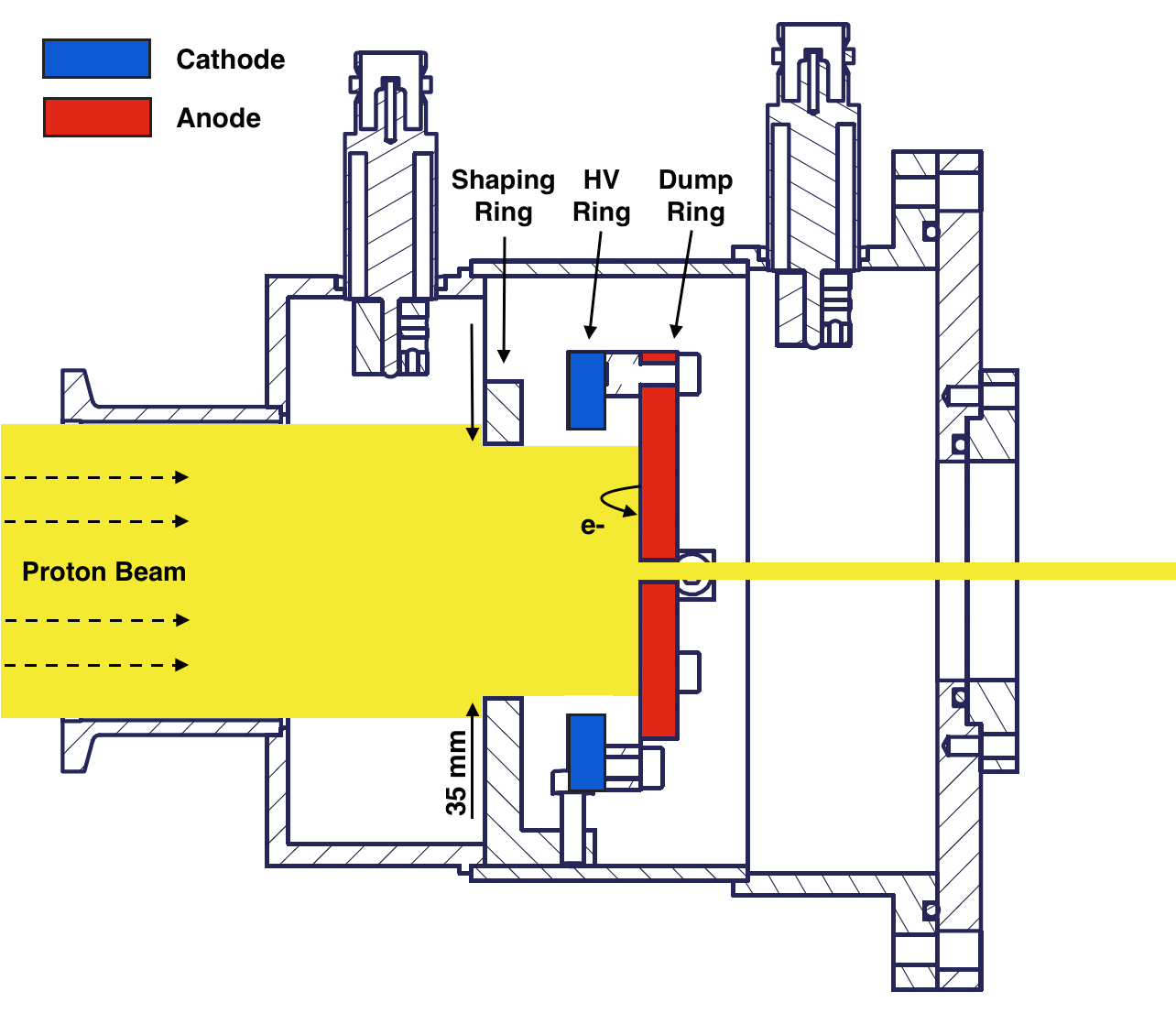}
  \caption{Cross section of the collimator system. The proton beam coming from the cyclotron enters the collimators system through an opening circular window of $40\,\si{\milli\meter}$. The beam is reduced to 35\,$\si{\milli\meter}$ diameter by means of a shaping ring. At this stage a flat beam of $35\,\si{\milli\meter}$ diameter approaches the dump ring. This acts as a collimator, so that part of the beam is extracted and delivered to the device under test through an aluminum extraction window, $300\,\si{\micro\meter}$ thick. The rest of the beam is absorbed on the dump ring so that the current density is measured.
When hitting the dump rings, the protons produce secondary electrons at the aluminum surface. For this reason a high voltage ring is in between the shaping ring and the dump ring. A bias voltage of $-100\,\si{\volt}$ is applied, so that the electrons are pushed back to the dump ring.}
  \label{fig:collimator}
\end{figure}

The current density is measured with a dedicated collimator, developed at the University of Bern. The cross-section of such colimator is shown in Fig.\,\ref{fig:collimator}.  It consists of three aluminum rings: the shaping ring, the high-voltage ring, and the beam dump. Each of the rings is $5\,\si{\milli\meter}$ thick, this thickness was optimised considering that $18\,\MeV$ protons have a range of $1.6\,\si{\milli\meter}$ in aluminum.

The proton beam coming from the BTL enters the system of collimators through a $40\,\si{\milli\meter}$ diameter opening and is further shaped through the shaping ring of $35\,\si{\milli\meter}$ inner diameter. After the shaping ring, the proton beam hits a dump ring, which also acts as collimator, allowing the beam to be extracted via a dedicated opening. Several versions of the dump ring are available, so that the beam can be extracted with different transverse shapes. Finally the beam is extracted into air through a $300\,\si{\micro\meter}$ aluminum window.
The part of the beam absorbed by the beam dump is used for the current density measurement. The dump ring is electrically connected to a high precision ammeter (Keysight Ammeter b2985a\,\cite{keysight}), which has a precision of $0.01\,\si{\femto\ampere}$ in the current measurement.

When the protons hit the aluminum surface of the beam dump, they extract secondary electrons. This phenomenon can lead to an overestimation of the current up to  $\sim20\,\%$ and it has been documented in Ref.\,\cite{Cyclo_LC}. In order to prevent this to happen, the system is biased with high-voltage. In particular an intermediate high voltage ring is then placed between the shaping ring and the dump ring. The high-voltage ring has an inner diameter of $37.5\,\si{\milli\meter}$ and is biased at $-100\,\si{\volt}$, preventing the extraction of secondary electrons in the dump ring. The bias voltage is provided by an ammeter.


The entire system is operated in vacuum in order to prevent air ionisation. This is necessary to avoid the beam to produce electrons which would be collected at the dump affecting the current measurement. For this reason the system is kept at $\sim10^{-8}\,\si{\bar}$, the vacuum system is the one provided by the BTL. External services are provided with two feedthrough connections: one devoted to a high voltage supply, and the other to connect the ammeter to the beam dump layer. 


The current per unit area $j(t)$ is calculated accordling to Eq.\,\ref{eq:curr_dens}:

\begin{align}
j(t) &= \frac{i(t)}{A_{BD}} \cdot \Lambda \, \label{eq:curr_dens};\\
A_{BD} &= \pi \cdot r^2 - A_{open} = 9.62\,\si{\centi\meter}^2 - A_{open}\, \label{eq:area};\\
\Lambda &= \dfrac{\rho_{extraction}}{\rho_{dump}}\, \label{eq:lambda}.
\end{align}
The current measured at the dump ring $i(t)$ is divided by the enlightened surface area $A_{BD}$ and corrected for beam inhomogeneities.
The area $A_{BD}$ is calculated in Eq.\,\ref{eq:area} and it depends on $A_{open}$, which is the extraction area. A correction factor $\Lambda$ takes in account the measurement of the two-dimensional beam profile. It is calculated as in Eq.\,\ref{eq:lambda}, where $\rho_{extraction}$ is the beam density at the extraction area  and $\rho_{dump}$ the one at the dump ring surface.


\section{The beam stopping power}
\begin{figure}
\centering
		\includegraphics[width=0.8\textwidth]{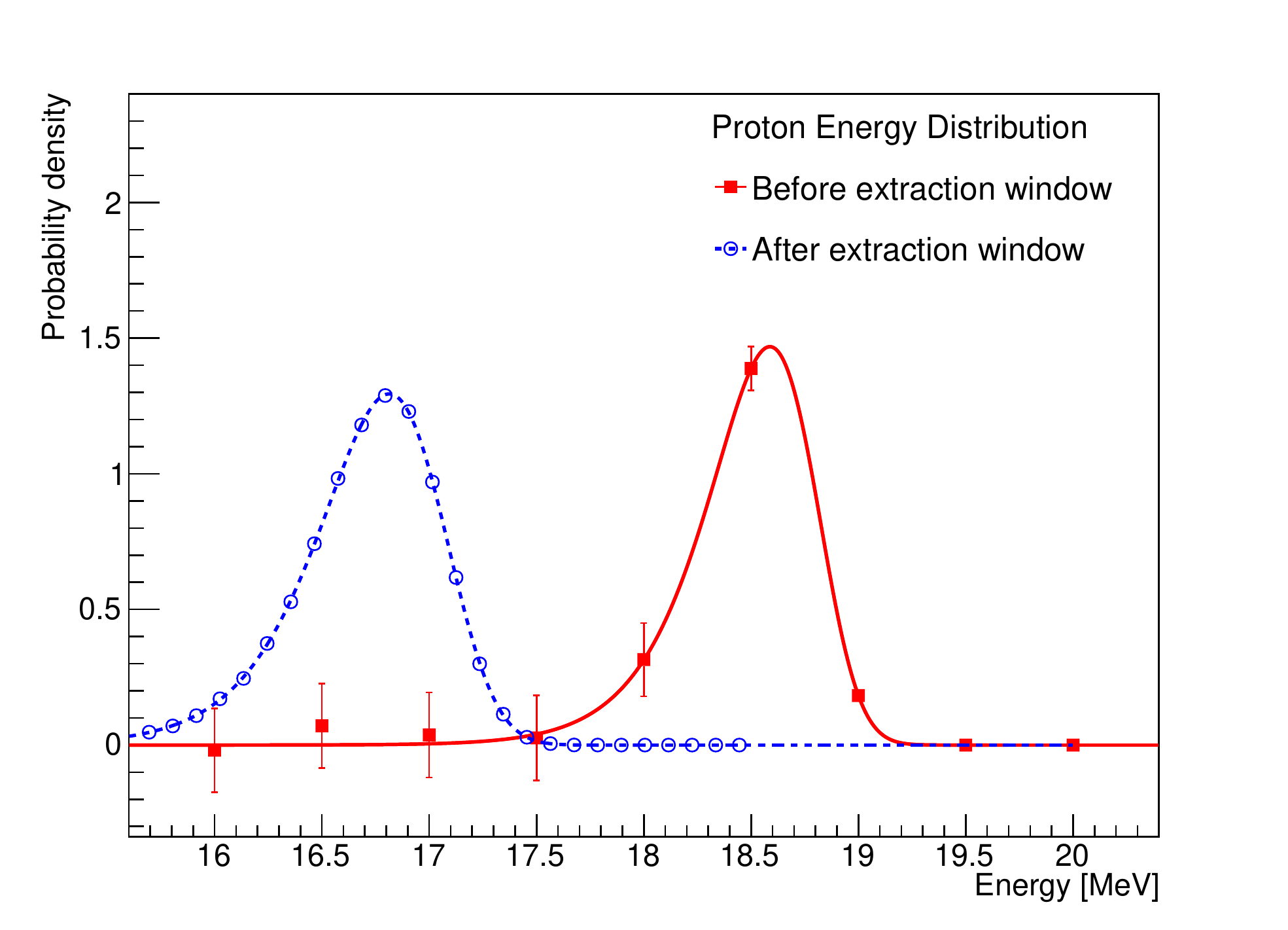}
		\caption{  Distribution of the proton energy. The square points are the measured beam profile fitted with the Verhulst function before the extraction window. The effect of $300\,\si{\micro\meter}$ aluminum extraction window has been simulated and it is shown by the open circle points. A dedicated Monte Carlo simulation was performed based on SRIM tables.}
		\label{fig:BeamEnergy}
\end{figure}
The current per unit area is not the only parameter to be taken in account for the dose measurement. Eq.\,\ref{eq:TID_est} shows that the stopping power $\dfrac{1}{\rho}\dfrac{dE}{dx}$ needs to be known. The stopping power depends on the beam energy and  the material of the device under irradiation.
The beam energy of the Bern Cyclotron was measured to be $18.3\pm0.4\,\MeV$\,\cite{BernEnergy}. The Bern Cyclotron beam energy distribution is shown by the square points in Fig.\,\ref{fig:BeamEnergy} which has been fitted with a Verlhust function shown by the red curve. The additional aluminum thickness of\,$300\si{\micro\meter}$ of the extraction window needs to be taken in account. A Monte Carlo simulation was performed, based on SRIM 2013\,\cite{SRIM} tables, to derive the beam energy probability density function after the extraction window, represented by the blue circles in Fig.\,\ref{fig:BeamEnergy}. A fit was performed with a Verhulst function convoluted with a gaussian, the mean energy was estimated to be $16.7\pm0.5\MeV$ after the extraction window.

\begin{table}
\caption{Stopping power for different material in case of proton beam of $16.7\,\MeV$. The values are obtained from the NIST database.}
\label{table:parameters}
\centering
\begin{tabular}{l|cc}
\toprule
Material  & $\dfrac{1}{\rho}\dfrac{dE}{\rho dx}_{ion}\,\big[\MeV \frac{\si{\centi\meter}^2}{\si{\gram}}\big]$ & $\dfrac{1}{\rho}\dfrac{dE}{\rho dx}_{nuc}\,\big[ \MeV\frac{\si{\centi\meter}^2}{\si{\gram}}\big]$\\
\midrule
Silicon & 23.25 & 0.011\\
Silicon dioxide & 24.31 & 0.012\\
Aluminum & 22.67 & 0.011 \\
Plastic scintillator (Vyniltoluene) & 30.12 & 0.015 \\
Mylar & 28.06 & 0.013 \\
Photographic emulsion & 18.30 & 0.009 \\
\bottomrule
\end{tabular}
\end{table}
Table\,\ref{table:parameters} shows the stopping power and the hardness factor for several materials considering a $16.7\,\MeV$ proton beam. The dose profile needs to be calculated depending on the material on the device under irradiation and its thickness. For example the protons will lose $\sim 0.55 \MeV$ in the first 100\,$\si{\micro\meter}$ in silicon.\\

\section{Performance and calibration of the irradiation setup}

\begin{figure}
\centering
		 \includegraphics[width=0.8\textwidth]{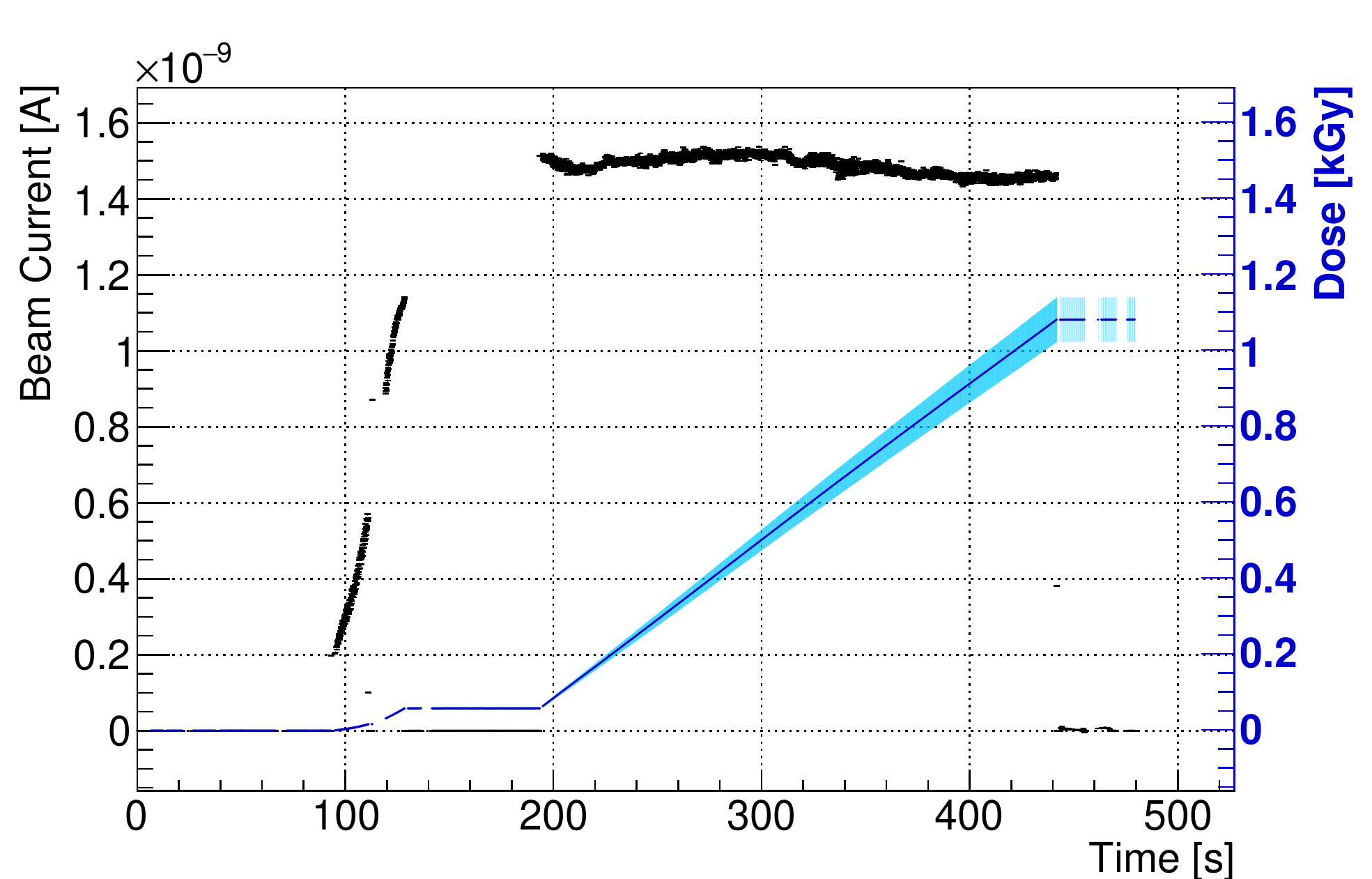}
  \caption{ Observed beam current during an irradiation with the estimated delivered dose given a test silicon device}
\label{fig:current_plot}
\end{figure}
Figure~\ref{fig:current_plot} shows the beam current intercepted by the dump ring and the dose measured during a typical irradiation taken as an example. The blue curve represents the dose, where the light blue error bar takes in consideration a $5\%$ uncertainty on the stopping power value and $500\,\si{\micro\meter}$ uncertainty on the acceptance radius of the dump ring. The experimental uncertainty on the current was measured to be negligible. The final uncertainty on the dose determination was estimated to be $6\,\%$ of the dose value. 
 A stability of approximatively $20\%$ of the beam current can be achieved by the operator with a manual adjustment of the beam. A software for online beam monitoring has been developed to provide the current and the dose measurement in real time. Data are updated every $250\,\si{\milli\second}$. Figure~\ref{fig:current_plot} shows data-taking example from irradiation that took place to evaluate the performance of the cyclotron. In the first 140\,$\si{\second}$ the beam current is increasing. This behavior was driven by the cyclotron operator, because the user wanted to observe the effects on the device due to an increasing beam current. After 200\,$\si{\second}$ the current is constant within 5\% and the beam parameters were kept constant until the end of the irradiation. The beam was shut down temporarily in 2 occasions because of user request.\\

\begin{figure}
\centering
\includegraphics[width=0.8\textwidth]{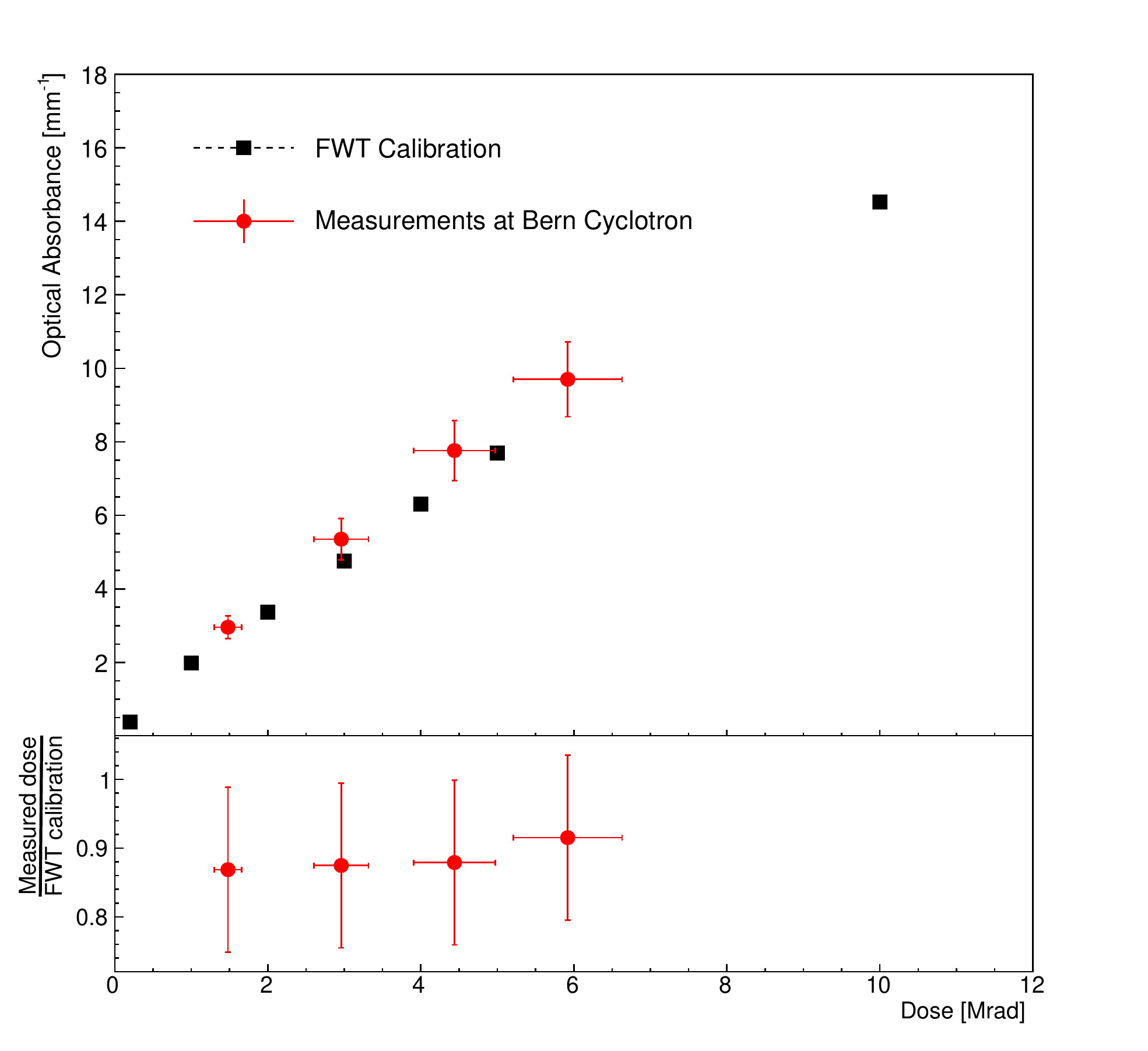}
\caption{Comparison between the film calibration provided by Far West Technologies and the dose measurement at the Bern Cyclotron. The two measurements techniques are compatible within 15\% as showed by the ratio in the bottom of the plot.}
\label{fig:film_cal}
\end{figure}

To check the assessment of the delivered dose, measurements were performed with radiochromic films. The film is positioned between the extraction window and the sample under irradiation. This measurement is available only at the end of the irradiation and is used as a cross-check of the delivered dose and the uniformity of the irradiation at the extraction of the beam. 
The radiochromic films used are FWT-60 made by Far West Technology (FWT)~\cite{FWT}. They are made of hexa(hydroxyethyl) aminotriphenylacetonitrile (HHEVC) dyes, and are available in different thickness: $10\,\si{\micro\meter}$ and $47\,\si{\micro\meter}$. Fig.\,\ref{fig:film_cal} shows the film calibration performed by FWT in the top plot with black squared points; the red round points show the measurement performed at the Bern Cyclotron where the dose is measured with the method described in Sections\,3 and 4. The measurements of the optical absorbance followed the prescription by FWT. The bottom plot shows the ratio of the dose measurement with Bern setup and the one provided by FWT at the same optical absorbance. This value is shown for the different dose measurements done with the Bern setup. In order to minimise the uncertainty on the optical absorbance several films were irradiated in indepedent measurements. The uncertainty on the measured optical absorbance takes in account an uncertainty of $5\,\si{\micro\meter}$ on the film thickness, as indicated by the FWT. The uncertainty of the dose measured in Bern follows the same consideration done for Fig.\,\ref{fig:current_plot} where the stopping power for a proton beam of $16.7\,\MeV$ in HHEVC was calculated to be $30.87\,\MeV\cdot\si{\centi\meter}^2\si{\gram}^{-1}$ with a dedicated SRIM simulation.   

\section{Conclusions}

An irradiation facility is in operation at the Bern Medical Cyclotron using the $18\,\si{\MeV}$ proton beam. Given the high precision control of the beam current and the online beam monitoring system, it is possible to have an online measurement of the total delivered dose, for both ionising and non-ionising phenomena. A dedicated system was developed for current monitoring during the irradiations, providing measurement of the total beam intensity and as well as the beam profile. A dose measurement technique was developed by the measurement of the beam current. The method was checked with pre-calibrated radio-chromic films and the two techniques have been found to be compatible within $15\%$. The maximum area of the extracted beam is a square of $2\times2\,\si{\centi\meter}^2$. Thanks to a downstream 2D-stage irradiation of samples of larger area is possible with multiple steps.
The irradiation facility is able to deliver from $0.1\,\frac{\si{\mega\radian}}{\si{\hour}}$ to $1\,\frac{\si{\giga\radian}}{\si{\hour}}$ thanks to adjustable beam current from  few $\si{\pico\ampere}$ to few hundreds $\si{\nano\ampere}$. 
A laboratory for sample characterisation is available on site for  further measurements of the irradiated samples. Two high-purity germanium detectors are available to check the isotope composition of the sample before and after irradiation. Electrical measurements can be performed with a precision ammeter, capable of measuring current down to to 0.01$\,\si{\femto\ampere}$. Further characterisation of the device under irradiation can be done with multi-channel low voltage power supplies available in-site. Cold storage of the irradiated samples is possible under request. A transition-current-technique setup is also available for characterisation of the bulk properties for solid-state detectors.
The Bern Cyclotron has, up to now, hosted different irradiation campaigns for high-energy physics experiments, such as ATLAS and LHCb, and as well as for astro-physics and medical-physics studies.
\bibliographystyle{BibStyle}
\bibliography{article_2.bib}
\end{document}